\documentclass[preprint,showpacs,showkeys,preprintnumbers,amsmath,amssymb]{revtex4}

\usepackage{graphicx}
\usepackage{epsfig}		
\usepackage{dcolumn}
\usepackage{bm}

\def\0{\mbox{\tiny $0$}}
\def\1{\mbox{\tiny $1$}}
\def\2{\mbox{\tiny $2$}}
\def\3{\mbox{\tiny $3$}}
\def\4{\mbox{\tiny $4$}}
\def\5{\mbox{\tiny $5$}}
\def\6{\mbox{\tiny $6$}}
\def\7{\mbox{\tiny $7$}}
\def\8{\mbox{\tiny $8$}}
\def\9{\mbox{\tiny $9$}}

\def\n{\mbox{\tiny $n$}}
\def\k{\mbox{\tiny $k$}}
\def\kk{\mbox{\small $k$}}

\def\f14{\mbox{\tiny $\frac{1}{4}$}}

\def\L{\mbox{\tiny $L$}}
\def\R{\mbox{\tiny $R$}}
\def\B{\mbox{\tiny $B$}}
\def\ii{\mbox{\tiny $i$}}

\def\z{\mbox{\tiny $z$}}
\def\s{\mbox{\tiny $s$}}
\def\r{\mbox{\tiny $r$}}

\def\x{\mbox{\tiny $x$}}
\def\y{\mbox{\tiny $y$}}

\def\j{\mbox{\tiny $j$}}
\def\mi{\mbox{\tiny $-$}}
\def\ig{\mbox{\tiny $=$}}
\def\pl{\mbox{\tiny $+$}}
\def\ppm{\mbox{\tiny $\pm$}}

\def\bb#1{\mbox{\footnotesize $(#1)$}}

\begin{document}


\title{Flavor coupled with chiral oscillations in the presence of an external
magnetic field}

\author{A. E. Bernardini}
\affiliation{Instituto de F\'{\i}sica Gleb Wataghin, UNICAMP,\\
PO Box 6165, 13083-970, Campinas, SP, Brasil.}
\email{alexeb@ifi.unicamp.br}

\date{\today}

\begin{abstract}
By reporting to the Dirac wave-packet prescription where it is formally assumed the {\em fermionic} nature of the particles, we shall demonstrate that chiral oscillations implicitly aggregated to the interference between positive and negative frequency components of mass-eigenstate wave-packets introduce some small modifications to the standard neutrino flavor conversion formula.
Assuming the correspondent spinorial solutions of a ``modified'' Dirac equation, we are specifically interested in quantifying flavor coupled with chiral oscillations for a {\em fermionic} Dirac-{\em type} particle (neutrino) non-minimally coupling with an external magnetic field {\boldmath$B$}. 
The viability of the intermediate wave-packet treatment becomes clear when we assume {\boldmath$B$} orthogonal/parallel to the direction of the propagating particle.
\end{abstract}

\maketitle

\section{Introduction}

Obtaining exact solutions of a generic class of Dirac wave equations \cite{Dir28,Esp99,Alh05,Ein81,Bak95}
becomes important since, for many times, the conceptual understanding of physics can only be brought about
by such solutions.
These solutions also correspond to valuable means for checking and improving models and numerical methods
for solving complicated physical problems.
In the context in which we intend to explore the Dirac formalism,
we can report about the Dirac wave-packet treatment which can be useful in keeping clear many of the
conceptual aspects of quantum oscillation phenomena that naturally arise in a relativistic spin
one-half particle theory. 
These studies have been paralleled by
much progress on the theoretical front of the quantum mechanics of neutrino oscillations,
both in phenomenological pursuit of a more refined flavor
conversion formula \cite{Giu98,Zra98,Ber05} which, sometimes, deserves a special attention,
and in efforts to give the theory a formal structure within quantum field formalism \cite{Bla95,Giu02B,Bla03}.
Under the point of view of a first quantized theory,
the flavor oscillation phenomena discussed in terms of
the {\em intermediate} wave-packet approach \cite{Kay81,Ber04} eliminates the most controversial
points rising up with the {\em standard} plane-wave formalism \cite{Kay89,Kay04}.
However, a common argument
against the {\em intermediate} wave-packet formalism is that oscillating
neutrinos are neither prepared nor observed \cite{Beu03}.
This point was clarified by Giunti
\cite{Giu02B} who suggested a solution by proposing an improved
version of the {\em intermediate} wave-packet model where the wave
packet of the oscillating particle is explicitly computed with
field-theoretical methods in terms of {\em external} wave-packets.
Since we intend to concentrate the discussion
on the Dirac equation properties for rigorously deriving a flavor/chiral conversion
formula for {\em fermionic} particles non-minimally coupling with an external magnetic
field, in this preliminary investigation,
we avoid the field theoretical methods in detriment to a clearer 
treatment with {\em intermediate} wave-packets
which commonly simplifies the understanding of physical aspects going with the oscillation
phenomena. 

From the experimental point of view, compelling evidences have continuously ratified
that neutrinos undergo flavor oscillations in vacuum or in matter.
For instance, we refer to the outstanding   
results of the Super-Kamiokande atmospheric neutrino experiment \cite{Fuk02},
in which a significant up-down asymmetry of the high-energy muon events was observed.
We also have the results of the SNO solar neutrino experiment \cite{Ahm02,Ban03} in which a direct evidence for the transition of the solar electron neutrinos into other flavors was obtained.
Finally, the KamLAND experiment \cite{Egu02} has confirmed
that the disappearance of solar electron neutrinos is mainly due to neutrino oscillations and not to other types of neutrino conversions \cite{Guz02,Bar02B}.
The experimental data could be completely interpreted and understood in terms
of three flavor quantum numbers, excepting by the LSND anomaly \cite{Ana98,Agu01,Ban03}
which permit us to speculate the existence of (at least) a fourth neutrino (flavor?) which
has to be inert.
In fact, the hypothesis of mixing between known neutrino species (electron, muon and tau)
and higher mass neutrinos including sterile neutrinos was naively studied in the literature \cite{Gel79,Yan79}.
The neutrino spin-flipping attributed to some dynamic external \cite{Oli90}
interacting process, which comes from the non-minimal coupling of a magnetic moment with an external electromagnetic
field \cite{Vol81}, was formerly supposed to be a relevant effect in the context of the solar-neutrino puzzle. 
As a consequence of a non-vanishing magnetic moment interacting with an
external electromagnetic field, left-handed neutrinos could change their helicity (to right-handed)\cite{Bar96}.
The effects on flavor oscillations due to external magnetic interactions in a kind of
chirality-preserving phenomenon were also studied \cite{Oli96} but they lack of a full detailed theoretical analysis.
In the standard model flavor-changing interactions, neutrinos with positive
chirality are decoupled from the neutrino absorbing charged weak currents \cite{DeL98}.
Consequently, such positive chirality neutrinos become sterile with respect to weak interactions.
Independently of any external electromagnetic field, since neutrinos are detected essentially via V-A
charged weak currents, the chiral oscillation mechanism by itself
may even explain the ``missing'' LSND data.
Despite the experimental circumstances not
being completely favorable to such an interpretation, which should be an additional motivation
to our theoretical calculations, the quantum transitions that produces a final flavor eigenstate
corresponding to an active-sterile quantum mixing is perfectly acceptable
from the theoretical point of view. 

To summarize, the main point to be considered in this manuscript concerns with accurately obtaining the complete flavor/chiral conversion
formula for fermionic particles, independently of its range of applicability which certainly shall be determined
in terms of more accurate experimental resolutions.
After such an introductory perspective, the first step of our study, which is presented in section II,
concerns with the {\em immediate} description of chiral oscillations 
in terms of the Hamiltonian dynamics ruled by the Lagrangian for the interaction between a fermionic field $\psi\bb{x}$ and an electromagnetic
field $F^{\mu\nu}\bb{x}$.
We investigate how the interactions with an external magnetic field can change the characteristics
of chiral oscillations which were previously obtained for free propagating particles in vacuum \cite{Ber05}.
By including the effect of chiral oscillation,
our final aim consists in verifying how the interaction with an external magnetic field 
can modify the neutrino flavor oscillation formula
which was previously established by means of what we call {\em scalar} prescription \cite{Ber04B}.
In section III, for treating the time-evolution of the spinorial
flavor-eigenstates, we shall take into account 
the chiral nature of charged weak currents and the time-evolution
of the chiral operator with Dirac wave-packets. 
To do it, we shall use the ``modified'' Dirac equation (with the above non-minimal coupling term) as the
evolution equation for the mass-eigenstates.
We just remark that we do not mind in developing the calculations concerning with 
analytic characteristics of the localization of each mass(flavor)-eigenstate since
all the modifications introduced by this prescription 
can be read independently of the wave-packet shapes.
However, we also dedicate an appendix to a small revision of the fundamental points of the wave-packet formalism
which, eventually, can be important for proceeding with subsequent analytic calculations.
Finally, we draw our conclusions in Section IV.

\section{Chiral oscillations in the presence of an external magnetic field}

In order to introduce the coupling with external magnetic fields, we observe that even presenting an electric charge neutrality, neutrinos can interact with a photon through loop (radiative) diagrams.
The Lagrangian for the interaction between a fermionic field $\psi\bb{x}$ and an electromagnetic
field written in terms of the field-strength tensor $F^{\mu\nu}\bb{x}= \partial^{\mu}A^{\nu}\bb{x} - \partial^{\nu}A^{\mu}\bb{x}$ is given by
\begin{equation}
\mathcal{L} = \frac{1}{2}\,\overline{\psi}\bb{x} \,\left[\sigma_{\mu\nu}\left(\mu\, F^{\mu\nu}\bb{x} - d \,\mathcal{F}^{\mu\nu}\bb{x}\right) + h.c. \right] \psi\bb{x}
\end{equation}
where $x = \bb{t, \mbox{\boldmath$x$}}$, $\sigma_{\mu\nu} = \frac{i}{2}[\gamma_{\mu},\gamma_{\nu}]$,
the {\em dual} field-strength tensor $\mathcal{F}^{\mu\nu}\bb{x}$ is given by $\mathcal{F}^{\mu\nu}\bb{x}= \frac{1}{2}\epsilon^{\mu\nu\lambda\delta} F^{\lambda\delta}\bb{x}$\footnote{$\epsilon^{\mu\nu\lambda\delta}$ is the totally fourth rank antisymmetric tensor.} and  
the coefficients $\mu$ and $d$ represent, respectively, the magnetic and the electric dipole moment which
establish the neutrino electromagnetic coupling.
One can notice that we have not discriminated the flavor/mass mixing elements in the above interacting Lagrangian since
we are indeed interested in the
physical observable dynamics ruled by the Hamiltonian 
\begin{eqnarray} 
\mathit{H} &=& \mbox{\boldmath$\alpha$}\cdot \mbox{\boldmath$p$} + \beta m 
		- \beta\left[\frac{\sigma_{\mu\nu}}{2}\left(\mu \,F^{\mu\nu}\bb{x}- d \mathcal{F}^{\mu\nu}\bb{x}\right) + h.c. \right]\nonumber\\
 		   &=& \mbox{\boldmath$\alpha$}\cdot \mbox{\boldmath$p$} + \beta \left[m - \mu\, \mbox{\boldmath$\Sigma$}\cdot \mbox{\boldmath$B$}\bb{x} - d \, \mbox{\boldmath$\Sigma$}\cdot \mbox{\boldmath$E$}\bb{x}\right] 
\label{01},~~
\end{eqnarray} 
where, in terms of the Dirac matrices, $\mbox{\boldmath$\alpha$} = \sum_{\k \ig \1}^{\3} \alpha_{\k}\hat{\kk} = \sum_{\k \ig \1}^{\3} \gamma_{\0}\gamma_{\k}\hat{\kk}$,
$\beta = \gamma_{\0}$, and $\mbox{\boldmath$B$}\bb{x}$ and $\mbox{\boldmath$E$}\bb{x}$ are respectively the magnetic and electric fields.
In fact, the Eq.~(\ref{01}) could be extended to an equivalent matrix representation with flavor and mass mixing elements 
where the diagonal (off-diagonal) elements described by $\mu_{\ii,\j}(m_{\ii,\j})$ and $d_{\ii,\j}(m_{\ii,\j})$,
where $i, \, j$ are mass indices, would be called diagonal (transition) moments.
In this context, for both Dirac and Majorana neutrinos, we could have transition amplitudes with non-vanishing 
magnetic and electric dipole moments \cite{Sch1,Mar1,Akh1}.
Otherwise, the CP invariance holds the diagonal electric dipole moments null \cite{Kim93}.
Specifically for Majorana neutrinos, 
it can be demonstrated that the diagonal magnetic and electric dipole moments vanish if $CPT$ invariance is assumed \cite{Sch1}.

Turning back to the simplifying example of diagonal moments and assuming CP and CPT invariance, 
we can restrict our analysis to the coupling with only an external magnetic field $\mbox{\boldmath$B$}\bb{x}$ by setting $d = 0$.
From this point, the expression for $\mu$ can be found from Feynman diagrams for magnetic moment corrections \cite{Kim93} and turns out to be
proportional to the neutrino mass (matrix),
\begin{equation} 
\mu = \frac{3\, e \,G}{8 \sqrt{2}\pi^{\2}} m = \frac{3\, m_e \,G}{4 \sqrt{2}\pi^{\2}}\, \mu_{\B} \, m_{\nu}
= 2.7 \times 10^{\mi \1\0}\,\mu_{\B}\,\frac{m_{\nu}}{m_N}
\end{equation} 
where $G$ is the Fermi constant and $m_{N}$ is the nucleon mass\footnote{We are using some results of the
standard $SU(2)_{L} \otimes U(1)_{Y}$ electroweak theory \cite{Gla61}.}.
In particular, for $m_{\nu}\approx 1 \, eV$, the magnetic moment introduced by the above formula is exceedingly small
to be detected or to affect astrophysical or physical processes.

Since we are interested in constructing the dynamics ruled by the Hamiltonian of Eq.~(\ref{01}),
we firstly observe that the free propagating momentum is not a conserved quantity,
\begin{equation} 
\frac{d~}{dt}\langle\mbox{\boldmath$p$}\rangle \,=\, i\langle\left[ \mathit{H} , \mbox{\boldmath$p$}\right]\rangle\,=\, \mu \langle \beta\, \mbox{\boldmath$\nabla$} \left(\mbox{\boldmath$\Sigma$}\cdot \mbox{\boldmath$B$}\bb{x}\right)\rangle
\label{02}
\end{equation} 
In the same way, the particle velocity given by
\begin{equation} 
\frac{d~}{dt}\langle\mbox{\boldmath$x$}\rangle \,=\, i\langle\left[ \mathit{H} , \mbox{\boldmath$x$}\right]\rangle\,=\,\langle\mbox{\boldmath$\alpha$}\rangle 
\label{03}
\end{equation}
comes out as a non-null value.
After solving the $\mbox{\boldmath$x$}\bb{t}$($\mbox{\boldmath$\alpha$}\bb{t}$) differential equation,
it is possible to observe that, in addition to a uniform motion, the fermionic particle
executes very rapid oscillations known as {\em zitterbewegung} \cite{Sak87}.
By following an analogous procedure for the Dirac chiral operator $\gamma_{\5}$,
newly recurring to the equation of the motion, it is possible to have
the chirality and the helicity dynamics respectively given by
\begin{eqnarray} 
\frac{d~}{dt}\langle \gamma^5 \rangle &=& 2\,i \langle\gamma_{\0}\,\gamma_{\5} \left[m - \mu\, \mbox{\boldmath$\Sigma$}\cdot \mbox{\boldmath$B$}\bb{x}\right]\rangle
\label{04}
\end{eqnarray} 
and
\begin{eqnarray} 
\frac{d~}{dt}\langle h \rangle &=& \frac{1}{2}\,\mu\,\langle \gamma_{\0} \left[(\mbox{\boldmath$\Sigma$}\cdot\mbox{\boldmath$\nabla$})(\mbox{\boldmath$\Sigma$}\cdot\mbox{\boldmath$B$}\bb{x})
+ 2(\mbox{\boldmath$\Sigma$}\times \mbox{\boldmath$B$}\bb{x})\cdot \mbox{\boldmath$p$}\right]\rangle
\label{05}
\end{eqnarray} 
where we have alternatively defined the particle helicity as 
the projection of the spin angular momentum onto the vector momentum,
$h = \frac{1}{2}\mbox{\boldmath$\Sigma$}\cdot{\mbox{\boldmath$p$}}$ (with
${\mbox{\boldmath$p$}}$ in place of $\hat{\mbox{\boldmath$p$}}$).
From Eqs.~(\ref{04}-\ref{05}) we can state that if a neutrino has an intrinsic magnetic moment and passes through a region filled by
an magnetic field, the neutrino helicity can flip in a completely different way from how chiral oscillations
evolve in time.
In the non-interacting case, it is possible to verify that the time-dependent averaged value of
the Dirac chiral operator $\gamma_{\5}$ has an oscillating behavior \cite{DeL98} very similar to
the rapid oscillations of the position.
The Eqs.~(\ref{04}-\ref{05}) can be reduced to the non-interacting case by setting $\mbox{\boldmath$B$}\bb{x} = 0$ so that
\begin{equation} 
\frac{d~}{dt}\langle h \rangle \,=\, i\langle\left[ \mathit{H} , h\right]\rangle\,=\,- \langle(\mbox{\boldmath$\alpha$} \times \mbox{\boldmath$p$})\cdot\hat{\mbox{\boldmath$p$}}\rangle \,=\ 0
\label{06}
\end{equation} 
and
\begin{equation} 
\frac{d~}{dt}\langle\gamma^{\5}\rangle \,=\, i\langle\left[ \mathit{H} , \gamma^{\5}\right]\rangle\,=\,2 \,i\,m \langle\gamma^{\0}\gamma^{\5}\rangle
\label{07}.
\end{equation} 
from which we confirm that the chiral operator $\gamma^{\5}$ is {\em not} a constant of the motion \cite{DeL98}.
The effective value of Eq.~(\ref{07}) appears only when both positive and negative frequencies are taken into account
to compose a Dirac wave-packet, i. e. 
the non-null expectation value of $\langle\gamma_{\0}\gamma_{\5}\rangle$ is revealed by the interference
between Dirac equation solutions with opposite sign frequencies.
The effective contribution due to this interference effect lead us to report to the Dirac wave-packet formalism
in order to quantify neutrino chiral oscillations in the presence of an external magnetic field.
 
Assuming the simplifying hypothesis of a uniform magnetic field  $\mbox{\boldmath$B$}$,
the physical implications of the non-minimal coupling with an external magnetic field can then be studied by means of the
eigenvalue problem expressed by the Hamiltonian equation
\begin{eqnarray} 
H\bb{\mbox{\boldmath$p$}} \, \varphi_{\n} = \,E_{\n}\bb{\mbox{\boldmath$p$}} \, \varphi_{\n}
 		   &=& \left\{\mbox{\boldmath$\alpha$}\cdot \mbox{\boldmath$p$} + \beta \left[m - \mu\, \mbox{\boldmath$\Sigma$}\cdot \mbox{\boldmath$B$}\right]\right\}\varphi_{\n}
\label{10},~~
\end{eqnarray} 
for which the explicit $4 \times 4$ matrix representation is given by
\begin{equation} 
H\bb{\mbox{\boldmath$p$}} \varphi_{\n} = \left[\begin{array}{cccc}
m - \mu B_{\z} & -\mu (B_{\x} - i B_{\y})& p_{\z} & p_{\x} - i p_{\y}\\
-\mu (B_{\x} + i B_{\y})& m + \mu B_{\z} & p_{\x} + i p_{\y} & -p_{\z}\\
p_{\z} & p_{\x} - i p_{\y}& - (m - \mu B_{\z}) & \mu (B_{\x} - i B_{\y})\\
 p_{\x} + i p_{\y} & -p_{\z} & \mu (B_{\x} + i B_{\y})& -(m + \mu B_{\z}) 
\end{array}\right] \varphi_{\n}
\label{10a}.~~
\end{equation}
The most general eigenvalue ($E_{\n}\bb{\mbox{\boldmath$p$}}$) solution of the above problem is given by 
\begin{eqnarray} 
E_{\n}\bb{\mbox{\boldmath$p$}} = \,  \pm E_{\s}\bb{\mbox{\boldmath$p$}} 
&=& \pm \sqrt{m^{\2} + \mbox{\boldmath$p$}^{\2} + \mbox{\boldmath$a$}^{\2} +\bb{\mi 1}^{\s}2
\sqrt{m^{\2}\mbox{\boldmath$a$}^{\2} + \bb{\mbox{\boldmath$p$} \times \mbox{\boldmath$a$}}^{\2}}}, ~~~~s\,=\, 1,\,2
\label{11},
\end{eqnarray} 
where we have denoted $E_{\n \ig \1,\2,\3,\4} = \pm E_{\s \ig \1,\2}$
and we have set $\mbox{\boldmath$a$} = \mu\, \mbox{\boldmath$B$}$.
The complete set of orthonormal eigenstates $\varphi_{\n}$ thus can be written in terms of the eigenfunctions $\mathcal{U}\bb{p_{\s}}$ with positive energy eigenvalues ($+ E_{\s}\bb{\mbox{\boldmath$p$}}$)
and the eigenfunctions $\mathcal{V}\bb{p_{\s}}$ with negative energy eigenvalues ($- E_{\s}\bb{\mbox{\boldmath$p$}}$),
\begin{eqnarray} 
\mathcal{U}\bb{p_{\s}} &=& N\bb{p_{\s}}\, \left\{\sqrt{\frac{A^{\mi}_{\s}}{A^{\pl}_{\s}}},\,\sqrt{\frac{\alpha^{\pl}_{\s}}{\alpha^{\mi}_{\s}}},\,\sqrt{\frac{A^{\mi}_{\s}\alpha^{\pl}_{\s}}{A^{\pl}_{\s}\alpha^{\mi}_{\s}}},\,-1\right\}^{\dagger}\nonumber\\
\mathcal{V}\bb{p_{\s}} &=& N\bb{p_{\s}}\, \left\{\sqrt{\frac{A^{\mi}_{\s}}{A^{\pl}_{\s}}},\,-\sqrt{\frac{\alpha^{\mi}_{\s}}{\alpha^{\pl}_{\s}}},\,-\sqrt{\frac{A^{\mi}_{\s}\alpha^{\mi}_{\s}}{A^{\pl}_{\s}\alpha^{\pl}_{\s}}},\,-1\right\}^{\dagger}
\label{12},
\end{eqnarray} 
where $p_{\s}$ is the relativistic {\em quadrimomentum}, $p_{\s} = (E_{\s}\bb{\mbox{\boldmath$p$}}, \mbox{\boldmath$p$})$,
$N\bb{p_{\s}}$ is the normalization constant and
\begin{equation} 
A^{\ppm}_{\s} = \Delta_{\s}^{\2}\bb{\mbox{\boldmath$p$}} \pm 2 m |\mbox{\boldmath$a$}|  - \mbox{\boldmath$a$}^{\2},~~~
\alpha^{\ppm}_{\s}= 2 E_{\s}\bb{\mbox{\boldmath$p$}} |\mbox{\boldmath$a$}| \pm (\Delta_{\s}^{\2}\bb{\mbox{\boldmath$p$}} + \mbox{\boldmath$a$}^{\2})
\nonumber
\end{equation}
with
\begin{equation}
\Delta_{\s}^{\2}\bb{\mbox{\boldmath$p$}} = 2\left(\mbox{\boldmath$a$}^{\2}
+\bb{\mi 1}^{\s}
\sqrt{m^{\2}\mbox{\boldmath$a$}^{\2} + \bb{\mbox{\boldmath$p$} \times \mbox{\boldmath$a$}}^{\2}}\right).
\end{equation}
We can observe that the above spinorial solutions are free of any additional constraint,
namely, at a given time $t$, they are independent functions of $\mbox{\boldmath$p$}$ and they do not represent chirality/helicity eigenstates.

In order to describe the above Hamiltonian dynamics for a generic observable $\mathcal{O}\bb{t}$
we can firstly seek a generic plane-wave decomposition as
\begin{eqnarray}
&& \exp{[- i(E_{\s}\bb{\mbox{\boldmath$p$}}\,t -\mbox{\boldmath$p$} \cdot \mbox{\boldmath$x$})]}\,\mathcal{U}\bb{p_{\s}},
~~~~\mbox{for positive frequencies and}\nonumber\\
&& \exp{[~ i(E_{\s}\bb{\mbox{\boldmath$p$}}\,t -\mbox{\boldmath$p$} \cdot \mbox{\boldmath$x$})]}\,\mathcal{V}\bb{p_{\s}},
 ~~~~\mbox{for negative frequencies},
\label{13}
\end{eqnarray}
so that the time-evolution of a plane-wave-packet $\psi\bb{t, \mbox{\boldmath$x$}}$ can be written as
\begin{eqnarray}
\psi\bb{t, \mbox{\boldmath$x$}}
&=& \int\hspace{-0.1 cm} \frac{d^{\3}\hspace{-0.1cm}\mbox{\boldmath$p$}}{(2\pi)^{\3}}
\sum_{\s \ig \1,\2}\{b\bb{p_{\s}}\mathcal{U}\bb{p_{\s}}\, \exp{[- i\,E_{\s}\bb{\mbox{\boldmath$p$}}\,t]}
\nonumber\\
&&~~~~~~~~~~~~~~~~
+ d^*\bb{\tilde{p}_{\s}}\mathcal{V}\bb{\tilde{p}_{\s}}\, \exp{[+i\,E_{\s}\bb{\mbox{\boldmath$p$}}\,t]}\}
\exp{[i \, \mbox{\boldmath$p$} \cdot \mbox{\boldmath$x$}]},
\label{14}
\end{eqnarray}
with $\tilde{p}_{\s} = (E_{\s},-\mbox{\boldmath$p$})$.
The Eq.~(\ref{14}) requires some extensive mathematical manipulations
for explicitly constructing the dynamics of an operator $\mathcal{O}\bb{t}$ like
\begin{equation}
\mathcal{O}\bb{t} = \int{d^{\3}\mbox{\boldmath$x$}\,\psi^{\dagger}\bb{t, \mbox{\boldmath$x$}}\,\mathcal{O}\,
\psi\bb{t, \mbox{\boldmath$x$}}}
\label{15}.
\end{equation}
If, however, the quoted observables like the chirality $\gamma^{\5}$,
the helicity $h$ or even the spin projection onto $\mbox{\boldmath$B$}$
commuted with the Hamiltonian $H$, we could reconfigure the above solutions to simpler ones. 
To illustrate this point, let us limit our analysis to very restrictive spatial configurations of $\mbox{\boldmath$B$}$ so that,
as a first attempt, we can calculate the observable expectation values which appear in Eq.~(\ref{04}).
Let us then assume that the magnetic field $\mbox{\boldmath$B$}$ is either orthogonal or parallel to the momentum $\mbox{\boldmath$p$}$.
For both of these cases the spinor eigenstates can then be decomposed into orthonormal bi-spinors as
\begin{equation}
\mathcal{U}\bb{p_{\s}} = N^{\pl}\bb{p_{\s}}\left[\begin{array}{r} \varphi^{\pl}\bb{p_{\s}}\\ \chi^{\pl}\bb{p_{\s}}\end{array}\right]
\end{equation}
and
\begin{equation} 
\mathcal{V}\bb{p_{\s}} = N^{\mi}\bb{p_{\s}}\left[\begin{array}{r} \varphi^{\mi}\bb{p_{\s}}\\ \chi^{\mi}\bb{p_{\s}}\end{array}\right]
\label{16}.
\end{equation}
Eventually, in order to simplify some subsequent calculations involving chiral oscillations, we could set 
$\varphi^{\ppm}_{\1,\2}$ and $\chi^{\ppm}_{\1,\2}$
as eigenstates of the spin projection operator $\mbox{\boldmath$\sigma$}\cdot\mbox{\boldmath$B$}$,
i. e. beside of being energy eigenstates, the general solutions $\mathcal{U}\bb{p_{\s}}$
and $\mathcal{V}\bb{p_{\s}}$ would become eigenstates of the operator $\mbox{\boldmath$\Sigma$}\cdot\mbox{\boldmath$B$}$ and, equivalently, of $\mbox{\boldmath$\Sigma$}\cdot\mbox{\boldmath$a$}$.

Now the Eq.(\ref{10}) can be decomposed into a pair of coupled equations like
\begin{eqnarray}
\left(\pm E_{\s} - m + \mbox{\boldmath$\sigma$}\cdot\mbox{\boldmath$a$} \right)\varphi^{\ppm}_{\s} &=& \pm \mbox{\boldmath$\sigma$}\cdot\mbox{\boldmath$p$}\,\chi^{\ppm}_{\s},\nonumber\\
\left(\pm E_{\s} + m - \mbox{\boldmath$\sigma$}\cdot\mbox{\boldmath$a$} \right)\chi^{\ppm}_{\s} &=& \pm \mbox{\boldmath$\sigma$}\cdot\mbox{\boldmath$p$}\,\varphi^{\ppm}_{\s},
\label{17}
\end{eqnarray}
where we have suppressed the $p_{\s}$ dependence.
By introducing the commuting relation 
$[\mbox{\boldmath$\sigma$}\cdot\mbox{\boldmath$p$},\,\mbox{\boldmath$\sigma$}\cdot\mbox{\boldmath$B$}] = 0$ 
which is derived when $\mbox{\boldmath$p$}\times\mbox{\boldmath$B$} = 0$,
the eigenspinor representation can be reduced to
\begin{equation}
\mathcal{U}\bb{p_{\s}} = \sqrt{\frac{E_{\s} + m_{\s}}{2E_{\s}}}
\left[\begin{array}{r} \varphi^{\pl}_{\s}\\ \frac{\mbox{\boldmath$\sigma$}\cdot\mbox{\boldmath$p$}}{E_{\s}+ m_{\s}}\,\varphi^{\pl}_{\s}\end{array}\right]\end{equation} and \begin{equation} 
\mathcal{V}\bb{p_{\s}} = \sqrt{\frac{E_{\s} + m_{\s}}{2E_{\s}}}
\left[\begin{array}{r} \frac{\mbox{\boldmath$\sigma$}\cdot\mbox{\boldmath$p$}}{E{\s}+ m_{\s}}\,\chi^{\mi}_{\s}\\ \chi^{\mi}_{\s}\end{array}\right]
\label{25},
\end{equation}
with $m_{\s} = m - \bb{\mi 1}^{\s}|\mbox{\boldmath$a$}|$
and the energy eigenvalues
\begin{equation}
\pm E_{\s} = \pm \sqrt{\mbox{\boldmath$p$}^{\2} + m_{\s}^{\2}}
\label{26}.
\end{equation}
In this case, the closure relations can be constructed in terms of 
\begin{eqnarray}
\sum_{\s \ig \1,\2}{\mathcal{U}\bb{p_{\s}}\otimes\mathcal{U}^{\dagger}\bb{p_{\s}}\gamma_{\0}}&=&
\sum_{\s \ig \1,\2}{\left\{\frac{\gamma_{\mu}p_{\s}^{\mu} + m_{\s}}{2 E_{\s}} 
\left[\frac{1-\bb{\mi 1}^{\s}\mbox{\boldmath$\Sigma$}\cdot\hat{\mbox{\boldmath$a$}}}{2}\right]\right\}}\nonumber\\
-\sum_{\s \ig \1,\2}{\mathcal{V}\bb{p_{\s}}\otimes\mathcal{V}^{\dagger}\bb{p_{\s}}\gamma_{\0}}&=&
 \sum_{\s \ig \1,\2}{\left\{\frac{-\gamma_{\mu}p_{\s}^{\mu} + m_{\s}}{2 E_{\s}} \left[\frac{1-\bb{\mi 1}^{\s}\mbox{\boldmath$\Sigma$}\cdot\hat{\mbox{\boldmath$a$}}}{2}\right]\right\}}.
\label{27}
\end{eqnarray}
Analogously, by introducing the anti-commuting relation
$\{\mbox{\boldmath$\sigma$}\cdot\mbox{\boldmath$p$},\,\mbox{\boldmath$\sigma$}\cdot\mbox{\boldmath$B$}\}$ when 
$\mbox{\boldmath$p$}\cdot\mbox{\boldmath$B$} = 0$, 
the eigenspinor representation can be reduced to
\begin{equation}
\mathcal{U}\bb{p_{\s}} = \sqrt{\frac{\varepsilon_{\0} + m}{2\varepsilon_{\0}}}
\left[\begin{array}{r} \varphi^{\pl}_{\s}\\ \frac{\mbox{\boldmath$\sigma$}\cdot\mbox{\boldmath$p$}}{\varepsilon_{\0}+ m}\,\varphi^{\pl}_{\s}\end{array}\right]\end{equation} and \begin{equation} 
\mathcal{V}\bb{p_{\s}} = \sqrt{\frac{\varepsilon_{\0} + m}{2\varepsilon_{\0}}}
\left[\begin{array}{r} \frac{\mbox{\boldmath$\sigma$}\cdot\mbox{\boldmath$p$}}{\varepsilon_{\0}+ m}\,\chi^{\mi}_{\s}\\ \chi^{\mi}_{\s}\end{array}\right]
\label{21},
\end{equation}
with $\varepsilon_{\0} = \sqrt{\mbox{\boldmath$p$}^{\2} + m^{\2}}$
and the energy eigenvalues
\begin{equation}
\pm E_{\s} = \pm\left[\varepsilon_{\0} + \bb{\mi 1}^{\s}|\mbox{\boldmath$a$}|\right]
\label{22}.
\end{equation}
In this case, the closure relations can be constructed in terms of 
\begin{eqnarray}
\sum_{\s \ig \1,\2}{\mathcal{U}\bb{p_{\s}}\otimes\mathcal{U}^{\dagger}\bb{p_{\s}}\gamma_{\0}}&=&
\frac{\gamma_{\mu}p_{\0}^{\mu} + m}{2 \varepsilon_{\0}} \sum_{\s \ig \1,\2}
{\left[\frac{1-\bb{\mi 1}^{\s}\gamma_{\0}\mbox{\boldmath$\Sigma$}\cdot\hat{\mbox{\boldmath$a$}}}{2}\right]}\nonumber\\
-\sum_{\s \ig \1,\2}{\mathcal{V}\bb{p_{\s}}\otimes\mathcal{V}^{\dagger}\bb{p_{\s}}\gamma_{\0}}&=&
\frac{-\gamma_{\mu}p_{\0}^{\mu} + m}{2 \varepsilon_{\0}} \sum_{\s \ig \1,\2}
{\left[\frac{1-\bb{\mi 1}^{\s}\gamma_{\0}\mbox{\boldmath$\Sigma$}\cdot\hat{\mbox{\boldmath$a$}}}{2}\right]},
\label{23}
\end{eqnarray}
where $p_{\0} = (\varepsilon_{\0}, \mbox{\boldmath$p$})$.

Since we can set $\varphi^{\pl}_{\1,\2} \equiv \chi^{\mi}_{\1,\2}$
as the components of an orthonormal basis,
we can immediately deduce the orthogonality relations
\begin{eqnarray}
&&\mathcal{U}^{\dagger}\bb{p_{\s}} \, \mathcal{U}\bb{p_{\r}} = 
\mathcal{V}^{\dagger}\bb{p_{\s}} \, \mathcal{V}\bb{p_{\r}} = \delta_{\s\r},
\nonumber\\
&&\mathcal{U}^{\dagger}\bb{p_{\s}} \,\gamma_{\0}\, \mathcal{V}\bb{p_{\r}} = 
\mathcal{V}^{\dagger}\bb{p_{\s}} \,\gamma_{\0}\, \mathcal{U}\bb{p_{\r}} = 0
\label{20}
\end{eqnarray}
which are valid for both of the above cases.

Finally, the calculation of the expectation value of $\gamma_{\5}\bb{t}$ is substantially simplified when
we substitute the above closure relations into the wave-packet expression of Eq.~(\ref{14}).
To clarify this point, we suppose
the initial condition over $\psi\bb{t,\mbox{\boldmath$x$}}$ can be set in terms of the Fourier transform of the weight function written as 
\begin{equation}
\varphi\bb{\mbox{\boldmath$p$}-\mbox{\boldmath$p$}_{\ii}}\,w \,=\,
\sum_{\s \ig \1,\2}{\{b\bb{p_{\s}}\mathcal{U}\bb{p_{\s}} + d^*\bb{\tilde{p}{\s}}\mathcal{V}\bb{\tilde{p}_{\s}}\}}
\label{28}
\end{equation}
so that
\begin{equation}
\psi\bb{0, \mbox{\boldmath$x$}}
= \int\hspace{-0.1 cm} \frac{d^{\3}\hspace{-0.1cm}\mbox{\boldmath$p$}}{(2\pi)^{\3}}
\varphi\bb{\mbox{\boldmath$p$}-\mbox{\boldmath$p$}_{\ii}}\exp{[i \, \mbox{\boldmath$p$} \cdot \mbox{\boldmath$x$}]}
\,w
\label{29}
\end{equation}
where $w$ is some fixed normalized spinor.
By using the orthonormality properties of Eq.~(\ref{20}),
we find \cite{Zub80}
\begin{eqnarray}
b\bb{p_{\s}} &=& \varphi\bb{\mbox{\boldmath$p$}- \mbox{\boldmath$p$}_{\ii}} \, \mathcal{U}^{\dagger}\bb{p_{\s}} \, w, \nonumber\\
d^*\bb{\tilde{p}_{\s}} &=& \varphi\bb{\mbox{\boldmath$p$}- \mbox{\boldmath$p$}_{\ii}}\,\mathcal{V}^{\dagger}\bb{\tilde{p}_{\s}}\, w.
\label{30}
\end{eqnarray}
For {\em any} initial state $\psi\bb{0, \mbox{\boldmath$x$}}$ given by Eq.~(\ref{29}),
the negative frequency solution coefficient
$d^*\bb{\tilde{p}_{\s}}$ necessarily provides a non-null contribution to the time-evolving wave-packet.
This obliges us to take the complete set of Dirac equation solutions to construct a complete and correct wave-packet solution.
Only if we consider the initial spinor $w$ being a positive energy ($E_{\s}\bb{\mbox{\boldmath$p$}}$) and momentum 
$\mbox{\boldmath$p$}$ eigenstate, the contribution due to the negative frequency solutions 
$d^*\bb{\tilde{p}_{\s}}$ will become null and we will have a simple expression for the time-evolution of
any physical observable.
By substituting the closure relations of Eqs.~(\ref{27}) and (\ref{23}) into the time-evolution expression
for the above wave-packet, the Eq.(\ref{14}) can thus be rewritten as
\begin{eqnarray}
\hspace{-0.1 cm}\psi\bb{t, \mbox{\boldmath$x$}}
&\hspace{-0.1 cm}=&
\hspace{-0.1 cm}\int\hspace{-0.1 cm} \frac{d^{\3}\hspace{-0.1cm}\mbox{\boldmath$p$}}{(2\pi)^{\3}}
\varphi\bb{\mbox{\boldmath$p$}-\mbox{\boldmath$p$}_{\ii}}\exp{[i \, \mbox{\boldmath$p$} \cdot \mbox{\boldmath$x$}]}
\sum_{\s \ig \1,\2}\mbox{$\left\{\left[\cos{[E_{\s}\,t]} -i\frac{H_{\s}}{E_{\s}}\sin{[E_{\s}\,t]}\right]\left(\frac{1-(\mi 1)^{\s}\mbox{\boldmath$\Sigma$}\cdot\hat{\mbox{\boldmath$a$}}}{2}\right)\right\}$}
w~~
\label{14A}
\end{eqnarray}
for the first case where $E_{\s}$ is given by Eq.~(\ref{26}) and $H_{\s} = \mbox{\boldmath$\alpha$}\cdot \mbox{\boldmath$p$} + \gamma_{\0} m_{\s}$, or as
\begin{eqnarray}
\hspace{-0.1 cm}\psi\bb{t, \mbox{\boldmath$x$}}
&\hspace{-0.1 cm}=&
\hspace{-0.1 cm}\int\hspace{-0.1 cm} \frac{d^{\3}\hspace{-0.1cm}\mbox{\boldmath$p$}}{(2\pi)^{\3}}
\varphi\bb{\mbox{\boldmath$p$}-\mbox{\boldmath$p$}_{\ii}}\exp{[i \, \mbox{\boldmath$p$} \cdot \mbox{\boldmath$x$}]}
\sum_{\s \ig \1,\2}\mbox{$\left\{\left[\cos{[E_{\s}\,t]} -i\frac{H_{\0}}{\varepsilon_{\0}}\sin{[E_{\s}\,t]}\right]\left(\frac{1-(\mi 1)^{\s}\gamma_{\0}\mbox{\boldmath$\Sigma$}\cdot\hat{\mbox{\boldmath$a$}}}{2}\right)\right\}$}
w~~~~
\label{14B}
\end{eqnarray}
for the second case where $\varepsilon_{\0}$ is given by Eq.~(\ref{22}) and $H_{\0} = \mbox{\boldmath$\alpha$}\cdot \mbox{\boldmath$p$} + \gamma_{\0} m$.

Now we can proceed with the calculation of the time-evolution of $\gamma_{\5}\bb{t}$ described by Eq.~(\ref{04})
assuming the simplifying hypothesis of a uniform magnetic field $\mbox{\boldmath$B$}$.
Once we have assumed the neutrino electroweak interactions at the source and detector are ({\em left}) chiral
$\left(\overline{\psi} \gamma^{\mu}(1 - \gamma^{\5})\psi W_{\mu}\right)$
only the component with negative chirality contributes to the propagation. 
It was already demonstrated that, in vacuum, chiral oscillations can introduce very small modifications
to the neutrino conversion formula \cite{DeL98,Ber05}.
The probability of a neutrino produced as a negative chiral eigenstate be detected after a time $t$ can be summarized by
\begin{eqnarray}
P(\mbox{\boldmath$\nu_{\alpha,\L}$}\rightarrow\mbox{\boldmath$\nu_{\alpha,\L}$};t) 
& = &
\int{d^{\3}\mbox{\boldmath$x$}\,\psi^{\dagger}\bb{t, \mbox{\boldmath$x$}}\,\frac{1 - \gamma_{\5}}{2}\,
\psi\bb{t, \mbox{\boldmath$x$}}} = \frac{1}{2}\left(1 - \langle\gamma_{\5}\rangle\bb{t}\right)
\label{15A}.
\end{eqnarray}
From this integral, it is readily seen that an initial $\mi 1$ chiral
mass-eigenstate will evolve with time changing its chirality.
By assuming the fermionic particle is created  at time $t=0$ as a $\mi 1$ chiral eigenstate ($\gamma_{\5} w = \mi w$),
in case of $[\mbox{\boldmath$\sigma$}\cdot\mbox{\boldmath$p$},\,\mbox{\boldmath$\sigma$}\cdot\mbox{\boldmath$B$}] = 0$
({\boldmath$B$} parallel to {\boldmath$p$}), we can write
\begin{eqnarray}
\langle\gamma_{\5}\rangle\bb{t} &=&  
\int\hspace{-0.1 cm} \frac{d^{\3}\hspace{-0.1cm}\mbox{\boldmath$p$}}{(2\pi)^{\3}}\varphi^{\2}\bb{\mbox{\boldmath$p$}-\mbox{\boldmath$p$}_{\ii}}\times\nonumber\\
&&~~
w^{\dagger}\sum_{\s \ig \1,\2}\mbox{$\left\{\left[\gamma_{\5}\cos^{\2}{[E_{\s}\,t]} +i \frac{[H_{\s},\,\gamma_{\5}]}{2E_{\s}}\sin{[2\,E_{\s}\,t]} + \frac{H_{\s}\,\gamma_{\5}\,H_{\s}}{E_{\s}^{\2}}\sin^{\2}{[E_{\s}\,t]}\right]\left(\frac{1-(\mi 1)^{\s}\mbox{\boldmath$\Sigma$}\cdot\hat{\mbox{\boldmath$a$}}}{2}\right)\right\}$}\,w\nonumber\\
&=&(\mi 1) \int\hspace{-0.1 cm} \frac{d^{\3}\hspace{-0.1cm}\mbox{\boldmath$p$}}{(2\pi)^{\3}}\varphi^{\2}\bb{\mbox{\boldmath$p$}-\mbox{\boldmath$p$}_{\ii}}
\sum_{\s \ig \1,\2}\mbox{$\left\{\left[\cos^{\2}{[E_{\s}\,t]} + \frac{\mbox{\boldmath$p$}^{\2} - m_{\s}^{\2}}{E_{\s}^{\2}}\sin^{\2}{[E_{\s}\,t]}\right]\,w^{\dagger}\left(\frac{1-(\mi 1)^{\s}\mbox{\boldmath$\Sigma$}\cdot\hat{\mbox{\boldmath$a$}}}{2}\right)w\right\}$}\nonumber\\
&=&(\mi 1) \int\hspace{-0.1 cm} \frac{d^{\3}\hspace{-0.1cm}\mbox{\boldmath$p$}}{(2\pi)^{\3}}\varphi^{\2}\bb{\mbox{\boldmath$p$}-\mbox{\boldmath$p$}_{\ii}}
\sum_{\s \ig \1,\2}\mbox{$\left\{\left[\frac{\mbox{\boldmath$p$}^{\2}}{E_{\s}^{\2}} + \frac{m_{\s}^{\2}}{E_{\s}^{\2}}\cos{[2\,E_{\s}\,t]}\right]\,w^{\dagger}\left(\frac{1-(\mi 1)^{\s}\mbox{\boldmath$\Sigma$}\cdot\hat{\mbox{\boldmath$a$}}}{2}\right)w\right\}$}
\label{15B},
\end{eqnarray}
where we have used the wave-packet expression of Eq.~(\ref{14A}) and, in the second passage, we have observed that
\begin{equation}
w^{\dagger} \gamma_{\5} w = \mi 1,~~~~  w^{\dagger} [H_{\s},\,\gamma_{\5}]w = 0 ~~~~\mbox{and}~~~~ H_{\s}\,\gamma_{\5}\,H_{\s} = \mbox{\boldmath$p$}^{\2} - m_{\s}^{\2}.
\label{15B1}
\end{equation}
The above expression can be reduced to a simpler one in the non-interacting case \cite{DeL98}.
Due to a residual interaction with the external magnetic field {\boldmath$B$} we can also observe chiral oscillations
in the ultra-relativistic limit.
However, from the phenomenological point of view, the coefficient of the oscillating term goes with $\frac{m_{\s}^{\2}}{E_{\s}^{\2}}$ which makes
chiral oscillations become not relevant for ultra-relativistic neutrinos \cite{Ber05,Ber05B}.
As a {\em toy model} illustration, by assuming a highly peaked momentum distribution centered around
a non-relativistic momentum $p_{\ii}\ll m_{\s}$, and with the wave-packet effects practically ignored, the chiral conversion formula
can be simplified and written as
\begin{equation}
P(\mbox{\boldmath$\nu_{\alpha,\L}$}\rightarrow\mbox{\boldmath$\nu_{\alpha,\L}$};t) 
 \approx  \frac{1}{2}\left(1 + \cos{[2\,m\,t]}\cos{[2\,|\mbox{\boldmath$a$}|\,t]}
 +\sin {[2\,m\,t]}\sin{[2\,|\mbox{\boldmath$a$}|\,t]}w^{\dagger}\mbox{\boldmath$\Sigma$}\cdot\hat{\mbox{\boldmath$a$}}w\right)
\label{15A222}
\end{equation}
where, in this case, all the oscillating terms come from the interference between positive and negative frequency solutions which compose the wave-packets.
Turning back to the case where
$\{\mbox{\boldmath$\sigma$}\cdot\mbox{\boldmath$p$},\,\mbox{\boldmath$\sigma$}\cdot\mbox{\boldmath$B$}\} = 0$
({\boldmath$B$} orthogonal to {\boldmath$p$}), we observe a phenomenologically more interesting result. 
By following a similar procedure with the mathematical manipulations, we can write
\begin{eqnarray}
\langle\gamma_{\5}\rangle\bb{t} &=&  
\int\hspace{-0.1 cm} \frac{d^{\3}\hspace{-0.1cm}\mbox{\boldmath$p$}}{(2\pi)^{\3}}\varphi^{\2}\bb{\mbox{\boldmath$p$}-\mbox{\boldmath$p$}_{\ii}}
w^{\dagger}\mbox{$\left\{\gamma_{\5}\cos{[E_{\1}\,t]}\cos{[E_{\2}\,t]} + \frac{H_{\0}\,\gamma_{\5}\,H_{\0}}{\varepsilon_{\0}^{\2}} \sin{[E_{\1}\,t]}\sin{[E_{\2}\,t]} +\right.$}\nonumber\\
&&~~~~~~~~~~~~~~~~~~~~~~~~~\left.\mbox{$\frac{i}{2}\left[[H_{\0},\,\gamma_{\5}]\sin{[E_{\1}+E_{\2}]} - \{H_{\0},\,\gamma_{\5}\}\gamma_{\0}\mbox{\boldmath$\Sigma$}\cdot\hat{\mbox{\boldmath$a$}}\sin{[E_{\1}-E_{\2}]} \right]$}\right\}w\nonumber\\
&=&(\mi 1) \int\hspace{-0.1 cm} \frac{d^{\3}\hspace{-0.1cm}\mbox{\boldmath$p$}}{(2\pi)^{\3}}\varphi^{\2}\bb{\mbox{\boldmath$p$}-\mbox{\boldmath$p$}_{\ii}}
\mbox{$\left\{\cos{[E_{\1}\,t]}\cos{[E_{\2}\,t]} + \frac{\mbox{\boldmath$p$}^{\2} - m^{\2}}{\varepsilon_{\0}^{\2}} \sin{[E_{\1}\,t]}\sin{[E_{\2}\,t]}\right\}$}\nonumber\\
&=&(\mi 1) \int\hspace{-0.1 cm} \frac{d^{\3}\hspace{-0.1cm}\mbox{\boldmath$p$}}{(2\pi)^{\3}}\varphi^{\2}\bb{\mbox{\boldmath$p$}-\mbox{\boldmath$p$}_{\ii}}
\mbox{$\left\{\frac{\mbox{\boldmath$p$}^{\2}}{\varepsilon_{\0}^{\2}}\cos{[2\,|\mbox{\boldmath$a$}|\,t]}+ \frac{ m^{\2}}{\varepsilon_{\0}^{\2}} \cos{[2\,\varepsilon_{\0}\,t]}\right\}$}
\label{15C},
\end{eqnarray}
where we have used the wave-packet expression of Eq.~(\ref{14B}) and, in addition to $w^{\dagger} \gamma_{\5} w = \mi 1$,
we have also observed that
$\{H_{\0},\,\gamma_{\5}\} = 2 \gamma_{\5}\mbox{\boldmath$\Sigma$}\cdot\hat{\mbox{\boldmath$p$}}$ and, subsequently, 
$w^{\dagger}\mbox{\boldmath$\Sigma$}\cdot\hat{\mbox{\boldmath$p$}}\gamma_{\0}\mbox{\boldmath$\Sigma$}\cdot\hat{\mbox{\boldmath$a$}} w = 0$.
Now, in addition to the non-interacting oscillating term $\frac{ m^{\2}}{\varepsilon_{\0}^{\2}} \cos{[2\,\varepsilon_{\0}\,t]}$, which comes from
the interference between positive and negative frequency solutions of the Dirac equation, we have
an extra term which comes from the interference between equal sign frequencies and, for very large time scales, can substantially change the oscillating results. 
In this case, it is interesting to observe that the ultra-relativistic limit of Eq.(\ref{15C}) leads to the following
expression for the chiral conversion formula,
\begin{equation}
P(\mbox{\boldmath$\nu_{\alpha,\L}$}\rightarrow\mbox{\boldmath$\nu_{\alpha,\L (\R)}$};t) 
 \approx  \frac{1}{2}\left(1 + (-) \cos{[2\,|\mbox{\boldmath$a$}|\,t]}\right)
\label{15A2}
\end{equation}
which, differently from chiral oscillations in vacuum, can be phenomenologically relevant.
Obviously, the Eq~(\ref{15A2}) is reproducing the consolidated results already attributed to neutrino spin-flipping
\cite{Kim93} where, only when we take ultra-relativistic limit ($m = 0$), the chirality quantum number can be approximated by the
helicity quantum number.
However, now it was accurately derived from the complete formalism with Dirac spinors.

\section{Flavor coupled to chiral oscillations in the presence of an external magnetic field}

After obtaining the time-evolution of the spinorial
mass-eigenstate wave-packets in the presence of an external magnetic field,
we have to immediately observe that we cannot arbitrarily overlook chiral oscillations while treating neutrino flavor oscillations.
Consequently, in order to quantify the interference of chiral oscillations over neutrino flavor conversion processes,
the chiral nature of charged weak currents and the time-evolution
of the chiral operator must be aggregated to the flavor oscillation formula.

Considering that the main aspects of oscillation phenomena can be understood by
studying a two flavor problem, we can introduce this simplifying condition for computing the
oscillation probabilities. Thus the time-evolution of flavor wave-packets can be described by
the state vector
\begin{eqnarray}
\Psi\bb{t, \mbox{\boldmath$x$}} &=& \psi_{\1}\bb{t, \mbox{\boldmath$x$}}\cos{\theta}\,\mbox{\boldmath$\nu_{\1}$} + \psi_{\2}\bb{t, \mbox{\boldmath$x$}}\sin{\theta}\,\mbox{\boldmath$\nu_{\2}$}\nonumber\\
          &=& \left[\psi_{\1}\bb{t, \mbox{\boldmath$x$}}\cos^{\2}{\theta} + \psi_{\2}\bb{t, \mbox{\boldmath$x$}}\sin^{\2}{\theta}\right]\,\mbox{\boldmath$\nu_\alpha$}+
		  \left[\psi_{\1}\bb{t, \mbox{\boldmath$x$}} - \psi_{\2}\bb{t, \mbox{\boldmath$x$}}\right]\cos{\theta}\sin{\theta}\,\mbox{\boldmath$\nu_\beta$}\nonumber\\
          &=& \psi_{\alpha}(t,\mbox{\boldmath$x$};\theta)\,\mbox{\boldmath$\nu_\alpha$} + \psi_{\beta}(t,\mbox{\boldmath$x$};\theta)\,\mbox{\boldmath$\nu_\beta$},
\label{0}
\end{eqnarray}
where {\boldmath$\nu_\alpha$} and {\boldmath$\nu_\beta$} are flavor-eigenstates,
{\boldmath$\nu_{\1}$} and {\boldmath$\nu_{\2}$} are mass-eigenstates and $\theta$ is the mixing angle.
The probability of finding a flavor state $\mbox{\boldmath$\nu_\beta$}$ at
the instant $t$ is equal to the integrated squared modulus of the
$\mbox{\boldmath$\nu_\beta$}$ coefficient,
\begin{equation}
P(\mbox{\boldmath$\nu_\alpha$}\rightarrow\mbox{\boldmath$\nu_\beta$};t)=
\int d^{\3}\mbox{\boldmath$x$} \,\left|\psi_{\beta}(t,\mbox{\boldmath$x$};\theta)\right|^{\2}
= \mbox{$\frac{\sin^{\2}{[2\theta]}}{2}$}\left\{\, 1 - \mbox{\sc Dfo}(t) \, \right\},
\label{1}
\end{equation}
where $\mbox{\sc Dfo}(t)$ represents the mass-eigenstate interference term explicitly computed in terms of
\begin{equation}
\mbox{\sc Dfo}(t) = \frac{1}{2} \int d^{\3}\mbox{\boldmath$x$}
\left[\psi^{\dagger}_{\1}\bb{t, \mbox{\boldmath$x$}} \psi_{\2}\bb{t, \mbox{\boldmath$x$}} + 
\psi^{\dagger}_{\2}\bb{t, \mbox{\boldmath$x$}}\psi_{\1}\bb{t, \mbox{\boldmath$x$}}\right]
\label{1AA}.
\end{equation}
Once we have assumed we have to take into account
the chiral conversion character for obtaining a complete description of the
flavor conversion mechanism, the most complete oscillation probability formula must be written as
\begin{equation}
P(\mbox{\boldmath$\nu_{\alpha,\L}$}\rightarrow\mbox{\boldmath$\nu_{\beta,\L}$};t) =
\mbox{$\frac{\sin^{\2}{[2\theta]}}{2}$}\left\{\, \mbox{\sc Dco}(t) - \mbox{\sc Dfco}(t) \, \right\}
\label{1AAA}
\end{equation}
where the results for $\mbox{\sc Dco}(t)$ corresponding exclusively to chiral
oscillations for each mass-eigenstate component
can be immediately reproduced from Eq.~(\ref{15A}) as
\begin{eqnarray}
\mbox{\sc Dco}(t) 
&=& \frac{1}{2} \int d^{\3}\mbox{\boldmath$x$}
\left[\psi^{\dagger}_{\1}\bb{t, \mbox{\boldmath$x$}} \frac{1 - \gamma_{\5}}{2}\psi_{\1}\bb{t, \mbox{\boldmath$x$}}
 + \psi^{\dagger}_{\2}\bb{t, \mbox{\boldmath$x$}}\frac{1 - \gamma_{\5}}{2}\psi_{\2}\bb{t, \mbox{\boldmath$x$}}\right] \nonumber\\
&=& \frac{1}{2}\left(1 - \frac{\langle\gamma_{\5}\rangle_{\1}\bb{t}+\langle\gamma_{\5}\rangle_{\2}\bb{t}}{2}\right)
\label{1AABC},
\end{eqnarray}
where the averaged values of $\gamma_{\5}$ can be, for instance, explicitly calculated in terms of the results of the Eqs.~(\ref{15B})
and (\ref{15C}) and, at the same time, the mixed flavor and chiral oscillation term can be given by 
\begin{equation}
\mbox{\sc Dfco}(t) = \frac{1}{2} \int d^{\3}\mbox{\boldmath$x$}
\left[\psi^{\dagger}_{\1}\bb{t, \mbox{\boldmath$x$}} \frac{1 - \gamma_{\5}}{2}\psi_{\2}\bb{t, \mbox{\boldmath$x$}}
 + \psi^{\dagger}_{\2}\bb{t, \mbox{\boldmath$x$}}\frac{1 - \gamma_{\5}}{2}\psi_{\1}\bb{t, \mbox{\boldmath$x$}}\right],
\label{1AABB}
\end{equation}
which deserves a more careful calculation.
We shall see then how we explicitly construct the complete oscillation formula containing both ``flavor-appearance'' (neutrinos of a
flavor not present in the original source) and ``chiral-disappearance'' (neutrinos of wrong chirality)
probabilities for both of the particular cases discriminated in section II.
The following results are obtained after some simple but extensive mathematical manipulations where,
again, we have imposed an initial constraint which establishes that
the normalizable mass-eigenstate wave functions $\psi_{\1,\2}\bb{t, \mbox{\boldmath$x$}}$
are created  at time $t=0$ as a negative chiral eigenstate ($w_{\1,\2}^{\dagger}\gamma_{\5}w_{\1,\2} = \mi 1$).
All the subsequent calculations do not depend on the gamma matrix representation.
In correspondence with the first case of the previous section, where
the propagating momentum {\boldmath$p$} is parallel to the magnetic field {\boldmath$B$},
we can write 
\small
\begin{eqnarray}
\mbox{\sc Dfco}(t) &=& \frac{1}{2}\int\frac{d^{\3} \mbox{\boldmath$p$}}{(2 \pi)^{\3}} \, \varphi^2\bb{\mbox{\boldmath$p$}}\sum_{\s\,\ig\,\1,\2}\left\{\,w^{\dagger}\left(\frac{1-(\mi 1)^{\s}\mbox{\boldmath$\Sigma$}\cdot\hat{\mbox{\boldmath$a$}}}{2}\right)w\right.\times \nonumber\\
&&\left.\left[
\left(1+\frac{\mbox{\boldmath$p$}^{\2}}{E^{(\1)}_{\s}E^{(\2)}_{\s}}\right)
\cos[\left(E^{(\1)}_{\s}-E^{(\2)}_{\s}\right) \, t] + 
\left(1-\frac{\mbox{\boldmath$p$}^{\2}}{E^{(\1)}_{\s}E^{(\2)}_{\s}}\right)
\cos[\left(E^{(\1)}_{\s}+E^{(\2)}_{\s}\right) \, t]\right]
\right\}
\label{2AABB},
\end{eqnarray}
\normalsize
where we have used the correspondence $\varphi^2\bb{\mbox{\boldmath$p$}} \equiv \varphi\bb{\mbox{\boldmath$p$}-\mbox{\boldmath$p$}_{\1}} \varphi\bb{\mbox{\boldmath$p$}-\mbox{\boldmath$p$}_{\2}}$
(see the discussion in the Appendix) and $E^{(\ii)}_{\s} = \sqrt{\mbox{\boldmath$p$}^{\2}+ m^{(\ii)}_{\s}}$ with $i = 1,2$ corresponding to the mass indices.
In fact, in the ultra-relativistic limit, and for not very strong magnetic fields ($|\mbox{\boldmath$a$}|\ll|\mbox{\boldmath$p$}|$),
the contribution due to the very rapid oscillations which come from the interference between negative and positive frequency components,
analogously to the case of {\em purely} chiral oscillations, does not introduce relevant modifications to the 
flavor conversion formula.
Otherwise, by taking the non-relativistic limit, with a momentum distribution sharply peaked around $|\mbox{\boldmath$p$}_{\ii}|\ll m_{\ii}$,
the complete oscillation probability formula can be written as
\begin{equation}
P(\mbox{\boldmath$\nu_{\alpha,\L}$}\rightarrow\mbox{\boldmath$\nu_{\beta,\L}$};t) =
\mbox{$\frac{\sin^{\2}{[2\theta]}}{4}$}
\sum_{\s\,\ig\,\1,\2}\left\{\,w^{\dagger}\left(\frac{1-(\mi 1)^{\s}\mbox{\boldmath$\Sigma$}\cdot\hat{\mbox{\boldmath$a$}}}{2}\right)w\,
\left(\cos{[m^{(\1)}_{\s} t]} - \cos{[m^{(\2)}_{\s} t]}\right)^{\2}\right\}
\label{1AAAA}
\end{equation}
which, despite not being phenomenologically verifiable,
introduces a completely different pattern for flavor/chiral oscillations.

Meanwhile, a more interesting interpretation is provided when we analyze the second case,
where the propagating momentum {\boldmath$p$} is orthogonal to the magnetic field {\boldmath$B$}.
Now, in the flavor/chiral oscillation formula, the effects of the external magnetic field can be discriminated from
the mass interference term, in the sense that we can write the complete expression for $\mbox{\sc Dfco}(t)$ as
\small
\begin{eqnarray}
\lefteqn{\mbox{\sc Dfco}(t) =\frac{1}{2} \int\frac{d^{\3} \mbox{\boldmath$p$}}{(2 \pi)^{\3}} \, \varphi^2\bb{\mbox{\boldmath$p$}}
\left\{\left(\cos[|\mbox{\boldmath$a$}^{(\1)}|\, t] \cos[|\mbox{\boldmath$a$}^{(\2)}|\, t] \right)\right.}\nonumber\\
&&\times
\left[\left(1+\frac{\mbox{\boldmath$p$}^{\2} + m^{(\1)}m^{(\2)}}{\varepsilon^{(\1)}_{\0}\varepsilon^{(\2)}_{\0}}\right)
\cos[\left(\varepsilon^{(\1)}_{\0}-\varepsilon^{(\2)}_{\0}\right) \, t] + 
\left(1-\frac{\mbox{\boldmath$p$}^{\2} + m^{(\1)}m^{(\2)}}{\varepsilon^{(\1)}_{\0}\varepsilon^{(\2)}_{\0}}\right)
\cos[\left(\varepsilon^{(\1)}_{\0}+\varepsilon^{(\2)}_{\0}\right) \, t]\right]\nonumber\\
&& ~~~~~~~~~~~~~~~~~~~~+
\frac{m^{(\1)}m^{(\2)}}{\varepsilon^{(\1)}_{\0}\varepsilon^{(\2)}_{\0}}\left[
\cos[\left(\varepsilon^{(\1)}_{\0}+\varepsilon^{(\2)}_{\0}\right) \, t]
\cos[\left(|\mbox{\boldmath$a$}^{(\1)}|-|\mbox{\boldmath$a$}^{(\2)}|\right)\, t]\right.
\nonumber\\
&&\left.\left.~~~~~~~~~~~~~~~~~~~~~~~~~~~~~~~~~~~~~~~~~~~~~~~~~~~~~ - 
\cos[\left(\varepsilon^{(\1)}_{\0}-\varepsilon^{(\2)}_{\0}\right) \, t]
\cos[\left(|\mbox{\boldmath$a$}^{(\1)}|+|\mbox{\boldmath$a$}^{(\2)}|\right)\, t] 
\right]
\right\},~~
\label{3AABB}
\end{eqnarray}
\normalsize
where we have used the correspondence of $\mbox{\boldmath$a$}^{(\ii)}$ and $\varepsilon^{(\ii)}$ with $m^{(\ii)}$.
Again, the ultra-relativistic limit reduces the impact of the modifications to residual effects which
are difficultly detectable by experiments.
By taking the non-relativistic limit and following the same procedure for obtaining the Eq.~(\ref{1AAAA}),
the complete oscillation probability formula can be written as
\begin{eqnarray}
P(\mbox{\boldmath$\nu_{\alpha,\L}$}\rightarrow\mbox{\boldmath$\nu_{\beta,\L}$};t) &=&
\mbox{$\frac{\sin^{\2}{[2\theta]}}{4}$}
\left(\cos^{\2}{[m^{(\1)} t]} + \cos^{\2}{[m^{(\2)} t]}\right.\nonumber\\
&&\left.~~~~~~~~~~ - 2\cos{[(|\mbox{\boldmath$a$}^{(\1)}|-|\mbox{\boldmath$a$}^{(\2)}|)t]} \cos{[m^{(\1)} t]}\cos{[m^{(\2)} t]} \right).
\label{1AAAAA}
\end{eqnarray}

Turning back to the starting point, if we had postulated a wave-packet
made up exclusively of positive frequency plane-wave solutions, 
the oscillation term which appear as a sum of mass-eigenstate energies would have vanished.
The new oscillations have very high frequencies.
Such a peculiar oscillating behavior is similar to the phenomenon referred to as {\em zitterbewegung}.
In atomic physics, the electron exhibits this violent quantum fluctuation in the position and
becomes sensitive to an effective potential which explains the Darwin term in the hydrogen atom \cite{Sak87}.
It reinforces the argument that, in constructing  Dirac wave
packets, we cannot simply forget the contributions due to negative
frequency components.

\section{Conclusions}

To quantify some subtle processes which accomplish the (standard) flavor oscillation phenomena \cite{Kay04}
and emerge via chiral oscillations of propagating neutrinos
non-minimally interacting with an external magnetic field {\boldmath$B$},
we have reported about some recent results on the study of flavor oscillation with Dirac wave-packets \cite{Ber04B}.
By taking into account the spinorial
form of neutrino wave functions and imposing an initial constraint
where a {\em pure} negative-chiral flavor-eigenstate is created at $t = 0$,
for a constant spinor $w$,
we have calculated the contribution to the wave-packet propagation
due to the interference between positive and negative frequency solutions
of the Dirac equation with non-minimal coupling and, finally,
we have obtained a complete expression for the oscillation probability.

It effectively represents the formally accurate way for deriving the expression for the neutrino
spin-flipping in magnetic fields related to chiral oscillations and coupled with a flavor conversion mechanism.
It is also to be noted that in the above discussion we have assumed
that neutrinos are Dirac particles, thus making the
positive-chiral component sterile. If the neutrino was a Majorana particle, it should not have a magnetic moment,
obviating the spin-flipping via magnetic field interactions
but still allowing the (vacuum) chiral conversion possibility via very rapid oscillations ({\em zitterbewegung}).

We have confirmed that the {\em fermionic} character of the particles modifies the standard oscillation
probability which is previously obtained by implicitly assuming a {\em scalar} nature of
the mass-eigenstates.
At the same time, it allows us to correctly determine the origin and the influence of chiral oscillations and spin-flipping
in the complete flavor conversion result.
Strictly speaking, we have obtained the term of very high oscillation frequency depending on the sum of energies
in the new oscillation probability formula which, in case of Dirac wave-packets, represents
some modifications that introduce correction factors which are negligible in the UR limit.
Our future perspectives concern with deriving the flavor coupled with chiral conversion expressions for neutrino
moving in the background matter by supposing that the magnitude of some experimentally (implicitly) observable
matter effects could be quantified (and eventually detected) in this Dirac wave-packet framework. 
In particular, under the point of view of a phenomenological analysis, we cannot discard the possibility
of existing larger magnetic momentum ($10^{\mi\1\2}\mu_{B}$)
for the electron/muon neutrinos deduced from some extensions of the minimal standard model \cite{Fuk1,Bab1}
which eventually can be implemented in a subsequent study.

We know, however, the necessity of a more sophisticated approach
is understood. In fact, the derivation of the oscillation formula should resort to field-theoretical methods\
which, meanwhile, are not very popular.
They are thought to be very complicated and the existing quantum field computations of the
oscillation formula do not agree in all respects \cite{Beu03}.
The  Blasone and Vitiello (BV) model \cite{Bla95,Bla03} to neutrino/particle mixing and oscillations
seems to be the most distinguished trying to this aim.
But still with Dirac wave-packets, the flavor conversion formula can be reproduced \cite{Ber04B}
with the same mathematical structure as those obtained in the BV model \cite{Bla95,Bla03}.
Moreover each new effect present in the oscillation formula can be separately quantified.

Just to summarize, we would not have not been honest if we had ignored the complete analysis of the general case comprised by
Eqs.~(\ref{10}-\ref{12}) where we had not yet assumed an arbitrary spatial configuration for the magnetic field.
Meanwhile, such a general case leads to the formal connection between quantum oscillation phenomena and
a very different field. It concerns with the curious fact that the above complete (general) expressions
for propagating wave-packets
do not satisfy the standard dispersion relations like $E^{\2} = m^{\2}+{\mbox{\boldmath$p$}}^{\2}$ excepting by the two
particular cases where $E_{\s}\bb{\mbox{\boldmath$p$}}^{\2} = m_{\s}^{\2}+ {\mbox{\boldmath$p$}}^{\2}$ for $\mbox{\boldmath$p$}\times\mbox{\boldmath$B$} = 0$
or $\varepsilon_{\0}^{\2} = m^{\2}+ {\mbox{\boldmath$p$}}^{\2}$ for $\mbox{\boldmath$p$}\cdot\mbox{\boldmath$B$} = 0$.
In principle, it could represent an inconvenient obstacle which forbids the extension of these restrictive cases to
a general one. However, we believe that it can also represent the starting point for discussing 
dispersion relations which can be incorporated into frameworks encoding the breakdown (or the validity) of Lorentz invariance. 

\begin{acknowledgments}
The author thanks FAPESP (PD 04/13770-0) for Financial Support.
\end{acknowledgments}

\appendix
\section{Flavor oscillations with wave-packets - Common points between {\em Scalar} and {\em Fermionic} prescriptions}

By reviewing the flavor conversion process for which we denominate a {\em scalar} prescription \cite{Ber04},
we notice that the time-evolution of flavor wave-packets can be described by
\begin{eqnarray}
\phi\bb{t, \mbox{\boldmath$x$}} &=& \phi_{\1}\bb{t, \mbox{\boldmath$x$}}\cos{\theta}\,\mbox{\boldmath$\nu_{\1}$} + \phi_{\2}\bb{t, \mbox{\boldmath$x$}}\sin{\theta}\,\mbox{\boldmath$\nu_{\2}$}\nonumber\\
          &=& \left[\phi_{\1}\bb{t, \mbox{\boldmath$x$}}\cos^{\2}{\theta} + \phi_{\2}\bb{t, \mbox{\boldmath$x$}}\sin^{\2}{\theta}\right]\,\mbox{\boldmath$\nu_\alpha$}+
		  \left[\phi_{\1}\bb{t, \mbox{\boldmath$x$}} - \phi_{\2}\bb{t, \mbox{\boldmath$x$}}\right]\cos{\theta}\sin{\theta}\,\mbox{\boldmath$\nu_\beta$}\nonumber\\
          &=& \phi_{\alpha}(t,\mbox{\boldmath$x$};\theta)\,\mbox{\boldmath$\nu_\alpha$} + \phi_{\beta}(t,\mbox{\boldmath$x$};\theta)\,\mbox{\boldmath$\nu_\beta$},
\label{0A}
\end{eqnarray}
where {\boldmath$\nu_\alpha$} and {\boldmath$\nu_\beta$} are flavor-eigenstates
and {\boldmath$\nu_{\1}$} and {\boldmath$\nu_{\2}$} are mass-eigenstates.
The probability of finding a flavor state $\mbox{\boldmath$\nu_\beta$}$ at
the instant $t$ is equal to the integrated squared modulus of the
$\mbox{\boldmath$\nu_\beta$}$ coefficient
\begin{equation}
P(\mbox{\boldmath$\nu_\alpha$}\rightarrow\mbox{\boldmath$\nu_\beta$};t)=
\int d^{\3}\mbox{\boldmath$x$} \,\left|\phi_{\beta}(t,\mbox{\boldmath$x$};\theta)\right|^{\2}
= \mbox{$\frac{\sin^{\2}{[2\theta]}}{2}$}\left\{\, 1 - \mbox{\sc Fo}(t) \, \right\},
\label{001}
\end{equation}
where $\mbox{\sc Fo}(t)$ represents the mass-eigenstate interference term given by
\begin{equation}
\mbox{\sc Fo}(t) = Re 
 \left[\, \int d^{\3}\mbox{\boldmath$x$}
\,\phi^{\dagger}_{\1}\bb{t,\mbox{\boldmath$x$}} \, \phi_{\2}\bb{t,\mbox{\boldmath$x$}} \, \right]\,
.
\label{2}
\end{equation}
As an illustrative example, we can consider {\em gaussian} wave-packets given at time $t = 0$ by
\begin{equation}
\phi_{\ii}(0,\mbox{\boldmath$x$}) = \left(\frac{2}{\pi a^{\2}}\right)^{ \frac{1}{4}} \exp{\left[- \frac{\mbox{\boldmath$x$}^{\2}}{a^{\2}}\right]} \exp{[i \mbox{\boldmath$p$}_{\ii} \, z]},
\label{3}
\end{equation}
where $s = 1,\, 2$.
The wave functions which describe their time-evolution are
\begin{eqnarray}
\phi_{\ii}(t,\mbox{\boldmath$x$}) =
\int\frac{d^{\3}\hspace{-0.1cm}\mbox{\boldmath$p$}}{(2 \pi)^{\3}} \,
\varphi(\mbox{\boldmath$p$} - \mbox{\boldmath$p$}_{\ii}) \exp{\left[-i\,E(\mbox{\boldmath$p$}, m_{\ii})\,t +i \, 
\mbox{\boldmath$p$}\cdot\mbox{\boldmath$x$}\right]},
\label{4}
\end{eqnarray}
where
$E(\mbox{\boldmath$p$}, m_{\ii}) = \left(\mbox{\boldmath$p$}^{\2} + m_{\ii}^{\2}\right)^{ \frac{1}{2}}$
and
\begin{equation}
\varphi(\mbox{\boldmath$p$} - \mbox{\boldmath$p$}_{\ii}) =  \left(2 \pi a^{\2} \right)^{ \frac{1}{4}} \exp{\left[- \frac{(\mbox{\boldmath$p$} - \mbox{\boldmath$p$}_{\ii})^{\2}\,a^{\2}}{4}\right]}.
\nonumber
\end{equation}
In order to obtain the oscillation probability, we can calculate the interference term $\mbox{\sc Fo}(t)$
by solving the following integral
\begin{eqnarray}
&&
\int \frac{d^{\3}\hspace{-0.1cm}\mbox{\boldmath$p$}}{(2 \pi)^{\3}} \,  \varphi(\mbox{\boldmath$p$} - \mbox{\boldmath$p$}_{ 1}) \varphi(\mbox{\boldmath$p$} - \mbox{\boldmath$p$}_{ 2})
\exp{[-i \, \Delta E(\mbox{\boldmath$p$}) \, t]} =
\nonumber\\ &&~~~~~~~~~~~~~~~~~~ 
\exp{\left[- \frac{(a \, \Delta{p})^{\2}}{8}\right]}
  \int\frac{d^{\3}\hspace{-0.1cm}\mbox{\boldmath$p$}}{(2 \pi)^{\3}}  \, \varphi^{\2}(\mbox{\boldmath$p$} -
\mbox{\boldmath$p$}_{\0})\exp{[-i \, \Delta E(\mbox{\boldmath$p$}) \, t]}, \label{6}
\end{eqnarray}
where we have changed the $z$-integration into a $\mbox{\boldmath$p$}$-integration
and introduced the quantities $\Delta |\mbox{\boldmath$p$}| = \mbox{\boldmath$p$}_{ \1} - \mbox{\boldmath$p$}_{ \2}
,\,\, \mbox{\boldmath$p$}_{\0} = \frac{1}{2}(\mbox{\boldmath$p$}_{ \1} + \mbox{\boldmath$p$}_{ \2})$
and $\Delta E(\mbox{\boldmath$p$}) = E(\mbox{\boldmath$p$}, m_{\1}) - E(\mbox{\boldmath$p$}, m_{\2})$.
The oscillation term is
bounded by the exponential function of $a \, \Delta |\mbox{\boldmath$p$}|$ at
any instant of time. Under this condition we could never observe a
{\em pure} flavor-eigenstate. Besides, oscillations are
considerably suppressed if $a \, \Delta |\mbox{\boldmath$p$}| > 1$. A necessary
condition to observe oscillations is that $a \, \Delta |\mbox{\boldmath$p$}| \ll 1$.
This constraint can also be expressed by $\delta |\mbox{\boldmath$p$}| \gg \Delta |\mbox{\boldmath$p$}|$
where $\delta |\mbox{\boldmath$p$}|$ is the momentum uncertainty of the particle. The
overlap between the momentum distributions is indeed relevant only
for $\delta |\mbox{\boldmath$p$}| \gg \Delta |\mbox{\boldmath$p$}|$. Consequently, without loss of
generality, we can assume
\begin{equation}
\mbox{\sc Fo}(t) = Re 
\left\{\int\frac{d^{\3}\hspace{-0.1cm}\mbox{\boldmath$p$}}{2
\pi}
 \, \varphi^{\2}(\mbox{\boldmath$p$} - \mbox{\boldmath$p$}_{\0})\exp{[-i \, \Delta E(\mbox{\boldmath$p$}) \, t]} \, \right\}
\label{9}.
\end{equation}

In literature, this equation is often obtained by assuming two
mass-eigenstate wave-packets described by the ``same'' momentum
distribution centered around the average momentum
$\bar{\mbox{\boldmath$p$}} = \mbox{\boldmath$p$}_{\0}$.
This simplifying hypothesis  also guarantees
{\em instantaneous} creation of a {\em pure} flavor
eigenstate {\boldmath$\nu_\alpha$} at $t = 0$.
In fact, for $\phi_{\1}(0,\mbox{\boldmath$x$})=\phi_{\2}(0,\mbox{\boldmath$x$})$ we get from Eq.~(\ref{0A})
\begin{equation}
\phi_{\alpha}(0,\mbox{\boldmath$x$},\theta) = \left(\frac{2}{\pi
a^{\2}}\right)^{\frac{1}{4}} \exp{\left[- \frac{\mbox{\boldmath$x$}^{\2}}{a^{\2}}\right]}
\exp{[i  \mbox{\boldmath$p$}_{\0} \cdot \mbox{\boldmath$x$}]}
\label{9B}
\end{equation}
and
$\phi_{\beta}(0,\mbox{\boldmath$x$},\theta) =0$.

The time-evolution of a spin one-half particle follows an analogous prescription.
Once we have introduced the {\em fermionic} character in the study of quantum oscillation phenomena,
we can use the Dirac equation as the evolution equation for the mass-eigenstates
represented by $\psi_{\ii}(t,\mbox{\boldmath$x$})$ so that the natural extension of Eq.~(\ref{9B}) reads
\begin{equation}
\psi_{\alpha}(0,\mbox{\boldmath$x$},\theta) = \phi_{\alpha}(0,\mbox{\boldmath$x$},\theta) \, w
\label{22BB}
\end{equation}
where we assume $w$ is the constant spinor which satisfies the normalization condition $w^{\dagger} w = 1$.

\end{document}